\documentclass[prl,aps,twocolumn]{revtex4-1}

\usepackage[pdftex]{graphicx}
\usepackage{amsbsy,amssymb,amsmath,bm,mathtools}

\usepackage{xspace}
\usepackage{bm}
\usepackage{color}

\usepackage{url}
\usepackage[section]{placeins}
\usepackage[colorlinks=true,linkcolor=blue,citecolor=blue]{hyperref}

\begin{document}


\title{Machine learning nonequilibrium electron forces for adiabatic spin dynamics}

\author{Puhan Zhang}

\author{Gia-Wei Chern}
\affiliation{Department of Physics, University of Virginia, Charlottesville, VA 22904, USA}

\date{\today}

\begin{abstract}
We present a generalized potential theory of nonequilibrium torques for the Landau-Lifshitz equation. The general formulation of exchange forces in terms of two potential energies allows for the implementation of accurate machine learning models for adiabatic spin dynamics  of out-of-equilibrium itinerant magnetic systems. To demonstrate our approach, we develop a deep-learning neural network that successfully learns the forces in a driven s-d model computed from the nonequilibrium Green's function method. We show that the Landau-Lifshitz dynamics simulations with forces predicted from the neural-net model accurately reproduce the voltage-driven domain-wall propagation. Our work opens a new avenue for multi-scale modeling of nonequilibrium dynamical phenomena in itinerant magnets and spintronics based on machine-learning models. 
\end{abstract}

\maketitle

In the past decade, machine learning (ML) techniques have greatly impacted many areas of industry and scientific research. The introduction of ML methods to the physical sciences has produced many fruitful results as well as opened several promising directions~\cite{carleo19,sarma19,bedolla21,carrasquilla17,nieuwenburg17,zhang17,schindler17,venderley18,carleo17,nomura17}. 
In particular, utilization of ML models as universal approximations for high-dimensional functions has significantly improved the efficiency of complex numerical simulations~\cite{rupp12,snyder12,brockherde17,schutt19,wang19,tsubaki20,burkle21,huang17,liu17,liu17b,nagai17,chen18}. 
Perhaps the most prominent and successful application in this direction is the ML potential for the prediction of energy and forces in quantum molecular dynamics (QMD) simulations~\cite{behler07,bartok10,botu17,li17,li15,smith17,zhang18,behler16,deringer19,mueller20,mcgibbon17,suwa19}. Contrary to classical MD methods that are based on empirical force fields, the atomic forces in QMD are computed by integrating out electrons on the fly as the atomic trajectories are generated~\cite{marx09}. Various many-body methods, notably the density functional theory (DFT), have been applied to solve the many-electron wave function for the force calculation of QMD. However, most of these electronic structure methods, even simple diagonalization or Hartree-Fock mean-field calculation, is computationally very expensive. The ML model offers a promising solution to this computational difficulty by accurately emulating the time-consuming many-body calculations,  thus offering the possibility of large-scale QMD simulations with the desired  quantum accuracy. 

The central idea behind the remarkable scalability of ML potentials is the principle of locality, or the nearsightedness, of electronic matter~\cite{kohn96,prodan05}, which, in the context of QMD simulations, assumes that the force acting on a given atom only depends on its immediate surroundings. A practical implementation of the ML potential based on this principle was demonstrated in the pioneer work of Behler and Parrinello~\cite{behler07}. In this approach, the total energy of the system is partitioned as $E = \sum_i \epsilon_i$, where $\epsilon_i$ is called the atomic energy and only depends on the local environment of the $i$-th atom~\cite{behler07,bartok10}. The complicated dependence of atomic energy on its neighborhood is provided by the ML model, which is trained on the condition that $E$ is given by the exact energy of the quasi-equilibrium electrons. 

Importantly, by focusing on the local energy $\epsilon_i$, which, as a scalar, is invariant under symmetry transformations such as rotations, the symmetry properties of the system can be easily incorporated into the ML model in such Behler-Parrinello (BP) schemes~\cite{behler07,bartok10}. The atomic forces are then obtained from derivatives of the ML energy:~$\mathbf F_i = -\partial E / \partial \mathbf R_i$, where $\mathbf R_i$ is the atomic position vector. This approach also ensures that the predicted forces are conservative, which is an important property for Born-Oppenheimber molecular dynamics simulations. The BP scheme has  been generalized to improve Monte Carlo simulations of lattice models in condensed matter physics~\cite{nagai20,ma19,zhang21a}. Notably, NN potentials based on the BP neural network have also been developed to enable large-scale Landau-Lifshitz dynamics simulations of correlated electron magnets~\cite{zhang20,zhang21}.

The fact that the atomic forces are conservative in the BP-type approach, however, also significantly limits its capability to represent forces due to highly nonequilibrium electrons, such as systems driven by an external voltage. This is because the energy $E$ is not a well defined concept in such open systems, and the resultant nonequilibrium forces often cannot be written as a derivative of an effective potential energy. A case in point is the current-induced force~\cite{lu12,todorov10,dundas09,diventra00} in, e.g. the molecular junctions, which has been shown to be nonconservative. Another important example is the spin-transfer torque~\cite{slonczewski96,berger96,brataas12,ralph08,salahuddin08} due to polarized electron current. Consequently, it is unclear how all the well-developed machinery of ML techniques for equilibrium QMD can be applied to model the dynamics of systems far from equilibrium.

In this paper, we propose a solution to this important problem in the context of quantum Landau-Lifshitz dynamics. We first show that the forces in a generalized Landau-Lifshitz equation can be expressed in terms of two potential energies. Applying the locality principle, a generalized BP neural network is developed to predict these two potential energies, from which the forces acting on spins can be obtained through automatic differentiation. We apply our ML framework to model the spin dynamics in a driven itinerant electron magnet. We further demonstrate the scalability of the trained NN potential by demonstrating large-scale LLG dynamics simulations of  voltage-driven domain-wall propagation.

We begin by developing a generalized force expression for the spin dynamics. The precessional dynamics of a magnetic system is described by the Landau-Lifshitz-Gilbert (LLG) equation: 
\begin{eqnarray}
	\label{eq:LL}
	\frac{\partial \mathbf S_i }{ \partial t }= - \gamma \mathbf S_i \times \mathbf H_i
	+\alpha  \mathbf S_i \times  \frac{\partial \mathbf S_i }{ \partial t } ,
\end{eqnarray} 
where $\gamma$ is the gyromagnetic ratio, $\mathbf H_i = - \partial E / \partial \mathbf S_i$ is the effective local magnetic field, which can be viewed as a force acting on the spins, and $E = E(\{\mathbf S_i \})$ is the energy of the system which is conserved by the first precessional term, also called the Landau-Lifshitz (LL) term, of the LLG equation. Dissipation effect is given by the second term, where $\alpha$ is an effective damping parameter.

Since the total energy $E$ is either conserved or, in the presence of dissipation, decreases with time, magnetization dynamics in an open system where energy can be pumped into spins from external sources is beyond the LL equation~(\ref{eq:LL}) with a conservative force. As noted above, the nonequilibrium electronic forces are often non-conservative and {\em cannot} be expressed as derivative of a single potential energy $E$. Consequently, the BP neural-net scheme cannot be naively applied to model the nonequilibrium forces. One possible solution is to develop ML models that can directly predict the nonconservative vector force $\mathbf H_i$ with input from the neighborhood~\cite{chmiela17,chmiela18}. However, besides the difficulty of including global symmetry with a vector output, ML force-field model without additional energy constraints is prone to overfitting and hence less accurate.

It is worth noting that although there is no energy conservation for out-of-equilibrium systems, it is still desirable to impose additional constraints on the ML model for force prediction. This this end, here we derive general expression of spin forces in terms of potential fields.  For simplicity, we consider classical spins of length $|\mathbf S_i| = 1$. The most general dynamical equation that preserves the spin length has the form $d \mathbf S_i / dt = - \mathbf S_i \times \mathbf V(\mathbf S_i)$, where $\mathbf V(\mathbf S)$ defines a vector field on a unit sphere $S^2$. Applying the Helmholtz-Hodge theorem for the case of the $S^2$ domain~\cite{adams62,swarztrauber81,fan18}, the vector field can be decomposed into radial, spheroidal, and toroidal components as:
\begin{eqnarray}
	\mathbf V(\mathbf S) =\mathbf S\, \mathcal{R}(\mathbf S) + \nabla_s\, \mathcal{E}(\mathbf S) +  \nabla_s \times \mathcal{G}(\mathbf S),
\end{eqnarray}
where $\mathcal{R}$, $\mathcal{E}$ and $\mathcal{G}$ are three scalar functions of $\mathbf S = (S^x, S^y, S^z)$,  the surface gradient and curl operators on a scalar function $f$ are defined as $\nabla_s f = \frac{\partial f }{\partial \mathbf S} - \mathbf S (\mathbf S\cdot \frac{\partial f }{\partial \mathbf S})$, and $\nabla_s \times f = \mathbf S \times \frac{\partial f }{\partial \mathbf S}$, respectively, and $\frac{\partial f }{\partial \mathbf S} = \sum_{\alpha = x, y, z} \frac{\partial f}{\partial S^\alpha}$ is the normal gradient  in three dimensions, without the restriction $|\mathbf S| = 1$. 

\begin{figure}
\includegraphics[width=1.0\columnwidth]{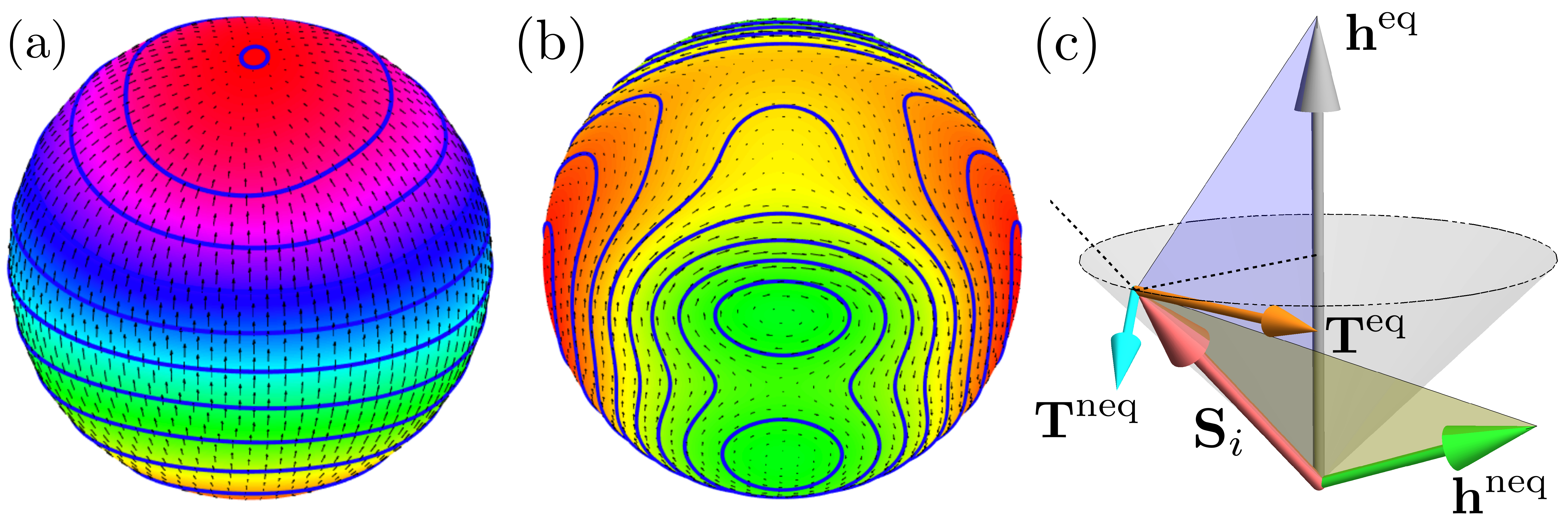}
\caption{A tangential vector field on a sphere can be decomposed into (a) the curl-free component $\nabla_s \, \mathcal{E}$ and (b) the divergence-free component $\nabla_s \times \mathcal{G}$. (c) shows the gradient field $\mathbf h^{\rm eq}_i = -\partial \mathcal{E}/\partial \mathbf S_i$, which can be viewed as a quasi-equilibrium exchange field, and the curl-field $\mathbf h^{\rm neq}_i = -\mathbf S_i \times \partial \mathcal{G} /\partial \mathbf S_i$, which corresponds to the nonequilibrium force, and their respective torques $\mathbf T^{\rm eq}_i = \mathbf h^{\rm eq}_i \times \mathbf S_i$ and $\mathbf T^{\rm neq}_i = \mathbf h^{\rm neq}_i \times \mathbf S_i$.}
\label{fig:decomposition} 
\end{figure}

Since the radial component, which is parallel to the spin direction, does not contribute to the torque, the radial function $\mathcal{R}$ is a pure gauge field. On the other hand, compared with the surface gradient $\nabla_s$, the normal gradient $\partial / \partial \mathbf S$ produces an additional radial component, which can also be gauged away. Consequently, the effective force can be expressed in terms of the two remaining scalar fields as
\begin{eqnarray}
	\label{eq:force}
	\mathbf H_i  = - \frac{\partial \mathcal{E}}{\partial \mathbf S_i} - \mathbf S_i \times \frac{\partial \mathcal{G}}{\partial \mathbf S_i} = \mathbf h_i^{\rm eq} + \mathbf h_i^{\rm neq}.
\end{eqnarray}
By analogy with the conservative force, the first term is called the quasi-equilibrium force. The second term which comes from the curl-field is denoted as the nonequilibrium force; see Fig.~\ref{fig:decomposition}.
The generalized LLG equation then reads
\begin{eqnarray}
	\label{eq:LL2}
	\frac{\partial \mathbf S_i }{ \partial t }= \gamma\, \mathbf S_i \times \frac{\partial \mathcal{E}}{\partial \mathbf S_i} + \gamma\, \mathbf S_i \times \left( \mathbf S_i \times \frac{\partial \mathcal{G}}{\partial \mathbf S_i} \right)
	+\alpha  \mathbf S_i \times  \frac{\partial \mathbf S_i }{ \partial t }, \qquad	
\end{eqnarray}
The first term is the conventional LL term in Eq.~(\ref{eq:LL}) with the system energy $E$ now replaced by the scalar potential~$\mathcal{E}$. Interestingly, the second term with $\mathcal{G} = -\lambda E$, where $\lambda > 0$ is a damping parameter and $E$ is the conservative energy, corresponds to a dissipation term introduced in LL's original work.  On the other hand, the nonequilibrium Slonczewski-Berger spin-torque~\cite{slonczewski96,berger96} can also be expressed by the second term in Eq.~(\ref{eq:LL2}) with the vector-field $\mathbf p_i = -\partial \mathcal{G} / \partial \mathbf S_i$ parallel to the magnetization of the fixed layer in a magnetic tunnel junction. These observations thus show that the energy non-conserving processes can be captured by the toroidal component $\mathcal{G}$ in Eq.~(\ref{eq:LL2}).

\begin{figure}
\includegraphics[width=1.0\columnwidth]{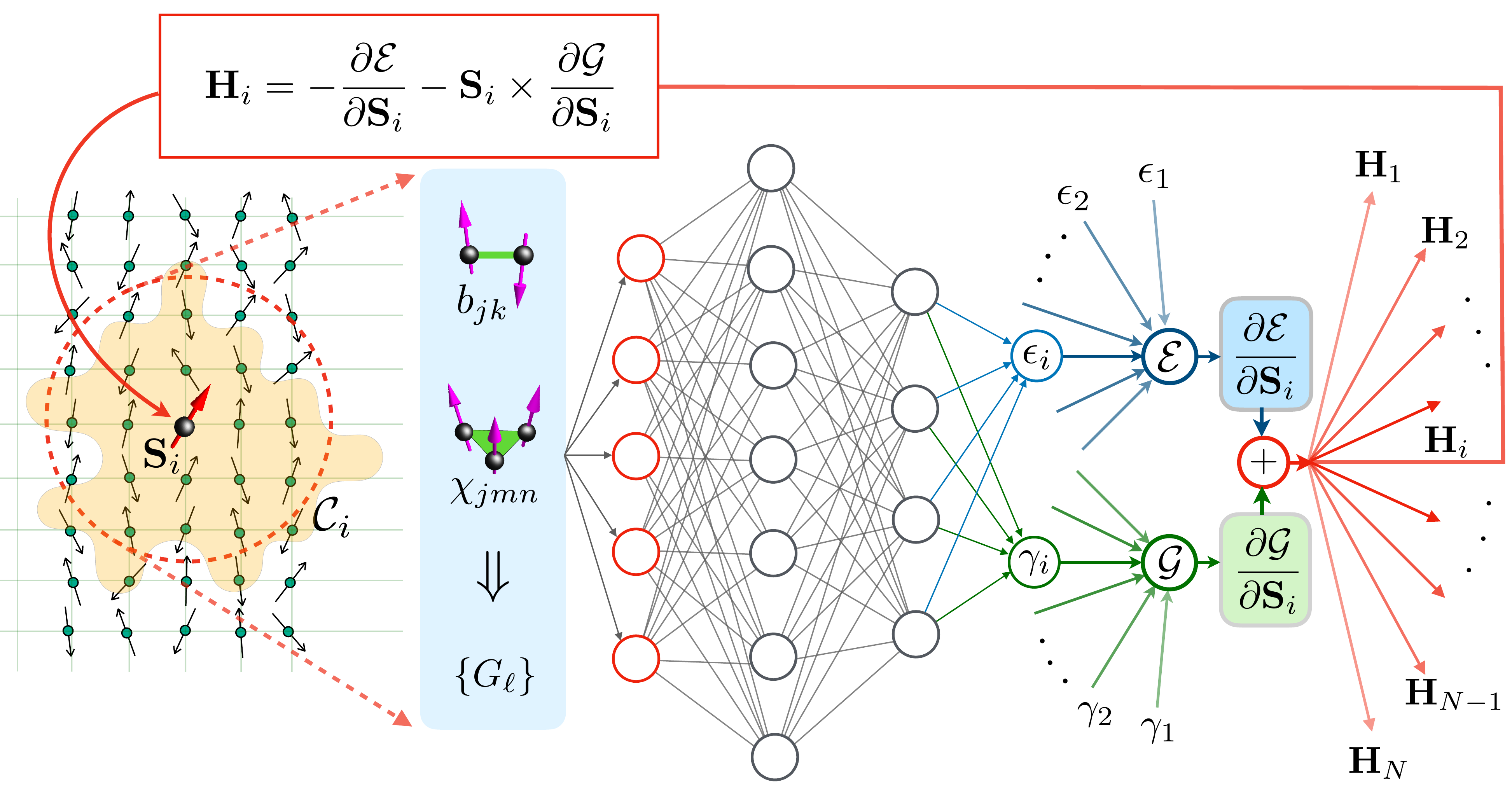}
\caption{Schematic diagram of neural-network (NN) potential model based on the generalized force expression~Eq.(\ref{eq:force}) for out-of-equilibrium itinerant spin system. }
\label{fig:ml-scheme} 
\end{figure}

Importantly, the generalized force formula in Eq.~(\ref{eq:force}) now allows us to generalize the BP-type NN model for the nonequilibrium electron forces. To this end, we partition the two potentials into local contributions, namely  $\mathcal{E} = \sum_i \epsilon_i$ and $\mathcal{G} = \sum_i \gamma_i$, based on the principle of locality~\cite{kohn96,prodan05}. These two local energies $\epsilon_i$ and $\gamma_i$ are assumed to depend only on the local magnetic environment $\mathcal{C}_i$ through two universal functions $\varepsilon(\cdot)$ and $\chi(\cdot)$ for a given electronic model. Consequently,  the dependence of the two potential energies on the spin configuration $\{\mathbf S_i\}$ can be expressed as
\begin{eqnarray}
	\label{eq:EG}
	\mathcal{E}(\{\mathbf S_j \}) =  \sum_i \varepsilon(\mathcal{C}_i), \quad \quad \mathcal{G}(\{\mathbf S_j \}) = \sum_i \chi(\mathcal{C}_i), \quad
\end{eqnarray}
In practice,  the magnetic environment $\mathcal{C}_i$ can be defined as the spin configuration within some cutoff radius $R_c$ from the $i$-th spin, i.e. $\mathcal{C}_i = \bigl\{ \mathbf S_j \, \big| \, |\mathbf r_j - \mathbf r_i| \le R_c \bigr\}$. The complex dependences of local energies on $\mathcal{C}_i$ are approximated by a deep-learning NN as shown in Fig.~\ref{fig:ml-scheme}.  Finally, the forces Eq.~(\ref{eq:force}) acting on spins are obtained from automatic differentiation of the two potentials.

The above ML framework is general and can be used to represent spin forces in any nonequilibrium electron systems. In the following, we apply it to model the forces computed from the nonequilibrium Green's functions (NEGF) method~\cite{meir92,jauho94,datta95,haug08,diventra08} for the well-studied itinerant s-d spin model~\cite{slonczewski96,berger96,brataas12,zhang04,yunoki98} under external voltage stress. We consider a square-lattice s-d system sandwiched by two electrodes in a capacitor structure. The total Hamiltonian is given by $\mathcal{H}_{\rm tot} = \mathcal{H}_{\rm sd} + \mathcal{H}_{\rm res}$, with
\begin{eqnarray}
	\label{eq:H_sd}
	\mathcal{H}_{\rm sd} = - t_{\rm nn} \sum_{\langle ij \rangle}\bigl( c^\dagger_{i\alpha} c^{\;}_{j\alpha} + \mbox{h.c.} \bigr)  
	 - J \sum_i  \mathbf S_i  \cdot  c^\dagger_{i\alpha} \bm\sigma^{\;}_{\alpha\beta} c^{\;}_{i\beta}, \qquad
\end{eqnarray}
and $\mathcal{H}_{\rm res}$ describes the electrodes and the reservoir bath degrees of freedom, as well as their coupling to the sd-model in the center. The effects of the reservoir fermions can be subsumed into a self-energy $\bm\Sigma^r$ in the retarded Green's function: $\mathbf G^r(\epsilon) = [\epsilon \mathbf I - \mathbf H_{\rm sd} - \bm \Sigma^r(\epsilon)]^{-1}$, where $\mathbf H_{\rm sd}$ is matrix representation of the sd Hamiltonian. Next, the lesser Green's function can be obtained using the Keldysh formula for quasi-steady electron states: $\mathbf G^{<}(\epsilon) = \mathbf G^r(\epsilon) \bm\Sigma^{<}(\epsilon) \mathbf G^a(\epsilon)$, where the lesser self-energy $\bm\Sigma^<$ is related to the $\bm\Sigma^{r}$ through the dissipation-fluctuation theorem. The electronic force on the spins, which is obtained from the generalized Hellmann-Feynman theorem, can then be computed from the lesser Green's function~\cite{stamenova05,salahuddin06,xie17,petrovic18,dolui20,chern21} 
\begin{eqnarray}
	\label{eq:H_NEGF}
	\mathbf H_i = - \biggl\langle   \frac{\partial \hat{\mathcal{H}}_{\rm sd}}{\partial \mathbf S_i }  \biggr\rangle = J \bm\sigma_{\beta\alpha} \int_{-\infty}^{+\infty} d\epsilon \, G^<_{i\alpha, i\beta}(\epsilon).
\end{eqnarray}
The exact forces computed from the NEGF method are used to train the NN model based on our generalized force formula. A magnetic descriptor developed in our previous work~\cite{zhang21} is employed to translate the local magnetic environment $\mathcal{C}_i$ into a set of feature variables $\{G_\ell\}$ that are invariant under symmetry operations of the system. In particular, in order to ensure the global spin rotation symmetry of the s-d model, the bond variables $b_{jk} = \mathbf S_j \cdot \mathbf S_k$ of spin-pairs  as well as the scalar chirality $\chi_{jkl} = \mathbf S_j \cdot \mathbf S_k \times \mathbf S_l$ of spin-triplets within the cutoff radius are used as the building blocks of the descriptor. These feature variables $\{G_\ell \}$ are fed into a fully connected NN, which in turn produces the two local energies $\epsilon_i$ and $\gamma_i$ associated with the $i$-th spin; see Fig.~\ref{fig:ml-scheme}. Applying the NN model to compute all the local energies, the two potential energies $\mathcal{E}$ and $\mathcal{G}$ are then obtained through Eq.~(\ref{eq:EG}). Finally, the automatic differentiation of these two potentials gives the local forces $\mathbf H_i$ acting on the spins.

\begin{figure}
\includegraphics[width=1.0\columnwidth]{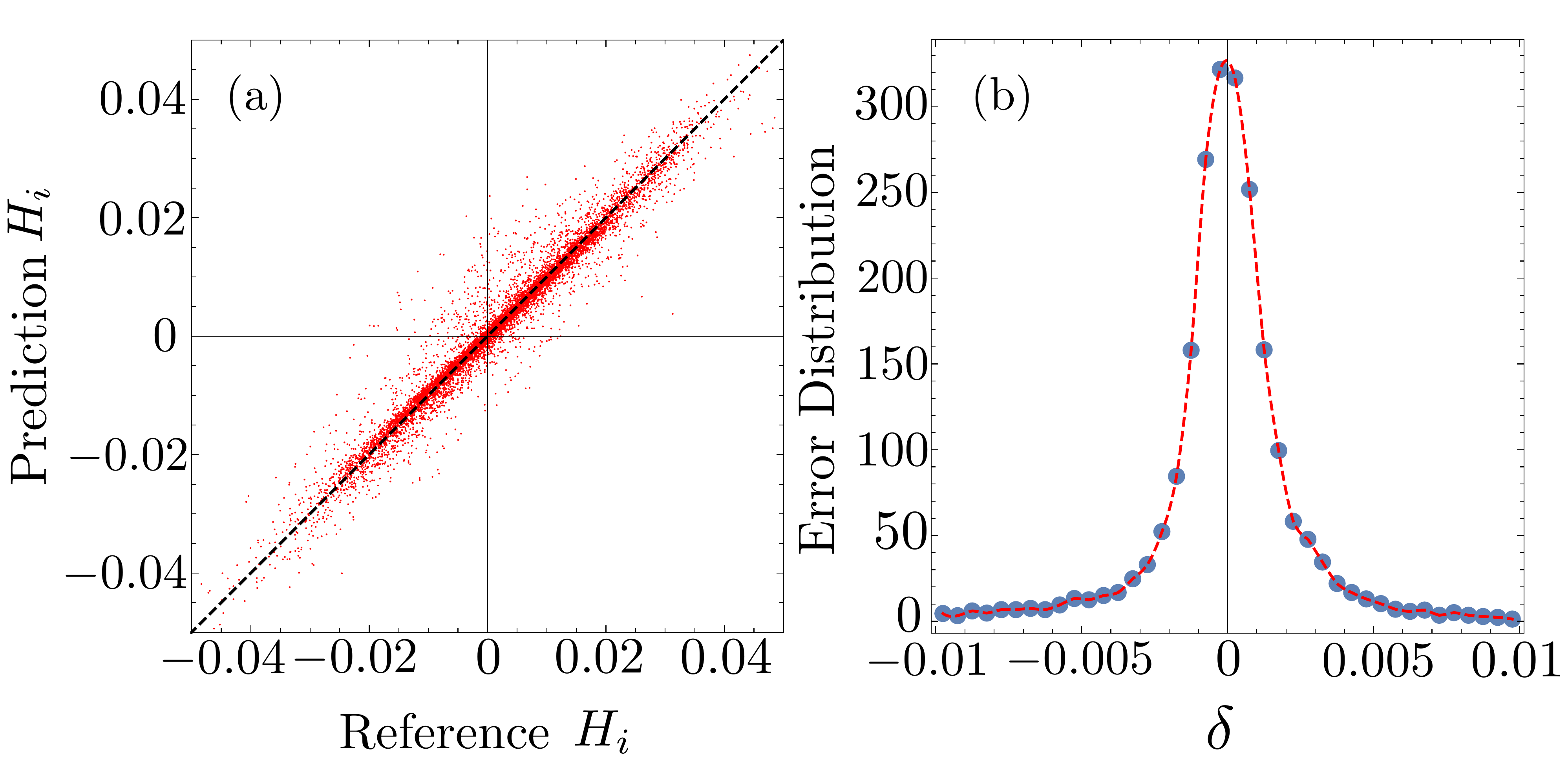}
\caption{(a) The ML predicted forces versus the exact solution from the NEGF calculation for the s-d model with exchange coupling $J = 3.8 t_{\rm nn}$. (b) Normalized distribution of the prediction error of the perpendicular components of the forces.}
\label{fig:forces} 
\end{figure}

A six-layer NN model is constructed and trained using PyTorch~\cite{paszke19} from the NEGF-LLG simulations on a $30\times 24$ lattice. A total of 3200 snapshots, each of which provides a total of $\sim 600$ force data,  are used for the training dataset. Contrary to the standard BP method where both forces and total energy are included in the training of the neural network, the loss function in our case is entirely given by the mean squared error of the perpendicular forces. Fig.~\ref{fig:forces}~(a) shows the componentwise torques $\mathbf S_i \times \mathbf H_i$ predicted from our trained NN model versus the exact results. An excellent MSE of $8.97 \times 10^{-6}$ is obtained from the trained NN model. The normalized distribution of the prediction error, shown in Fig.~\ref{fig:forces}(b), is characterized by a rather small standard deviation of~$\sigma = 0.0014$.

We next incorporate the NN potential model into the LLG dynamics for the simulation of the voltage-driven domain-wall propagation in the square-lattice s-d model.  The system is initially in an insulating antiferromagnetic (AFM) state with a spectral gap $\Delta E_g = 2 J$. An external voltage $V$ is applied to the two electrodes, which couple to the system at the left and right edges. When the chemical potential of one electrode is lowered to the eigen-energies of the in-gap edge modes, an instability towards the ferromagnetic (FM) ordering is triggered as electrons are drained from the edge of the system into the electrode. This instability leads to the nucleation of the FM domains at the edge of the sample.  The subsequent expansion of the FM domains transforms the system into the low-resistant metallic state.

\begin{figure}
\includegraphics[width=0.99\columnwidth]{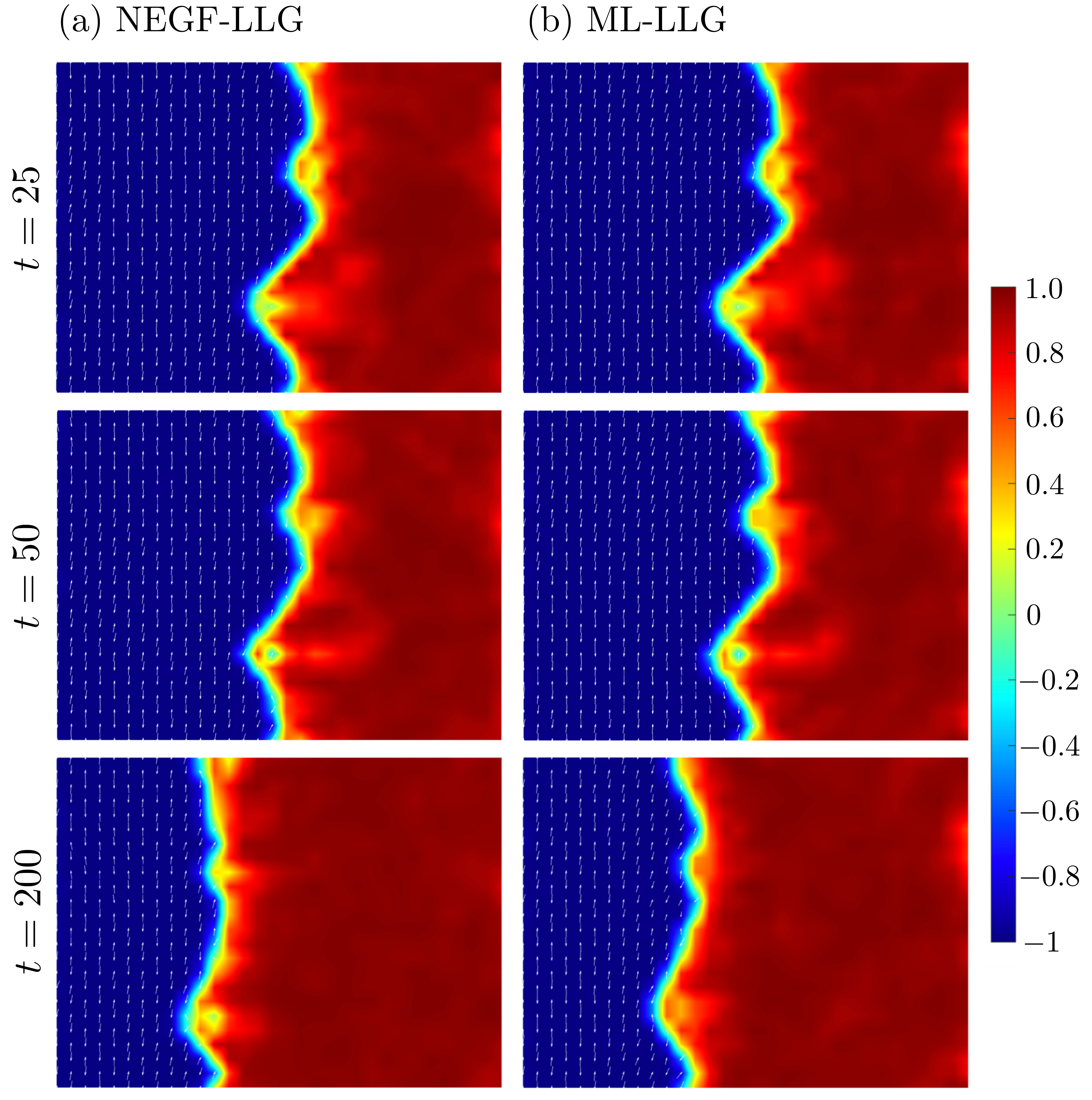}
\caption{Domain wall propagation in a s-d system driven by an external voltage $eV = 3.2 t_{\rm nn}$. Comparison between (left) NEGF-LLG simulations and (right) ML-LLG simulations. The lattice size is $30\times 24$.  The color bar indicates the local nearest-neighbor spin correlation $b_{ij} = \mathbf S_i \cdot \mathbf S_j$. Blue (red) area corresponds to AFM (FM) domains. The NEGF-LLG simulation was carried out at a low temperature of $T = 0.01\,t_{\rm nn}$, while the ML simulation was performed without Langevin noise.
\label{fig:domain}
}
\end{figure}

\begin{figure}
\includegraphics[width=0.99\columnwidth]{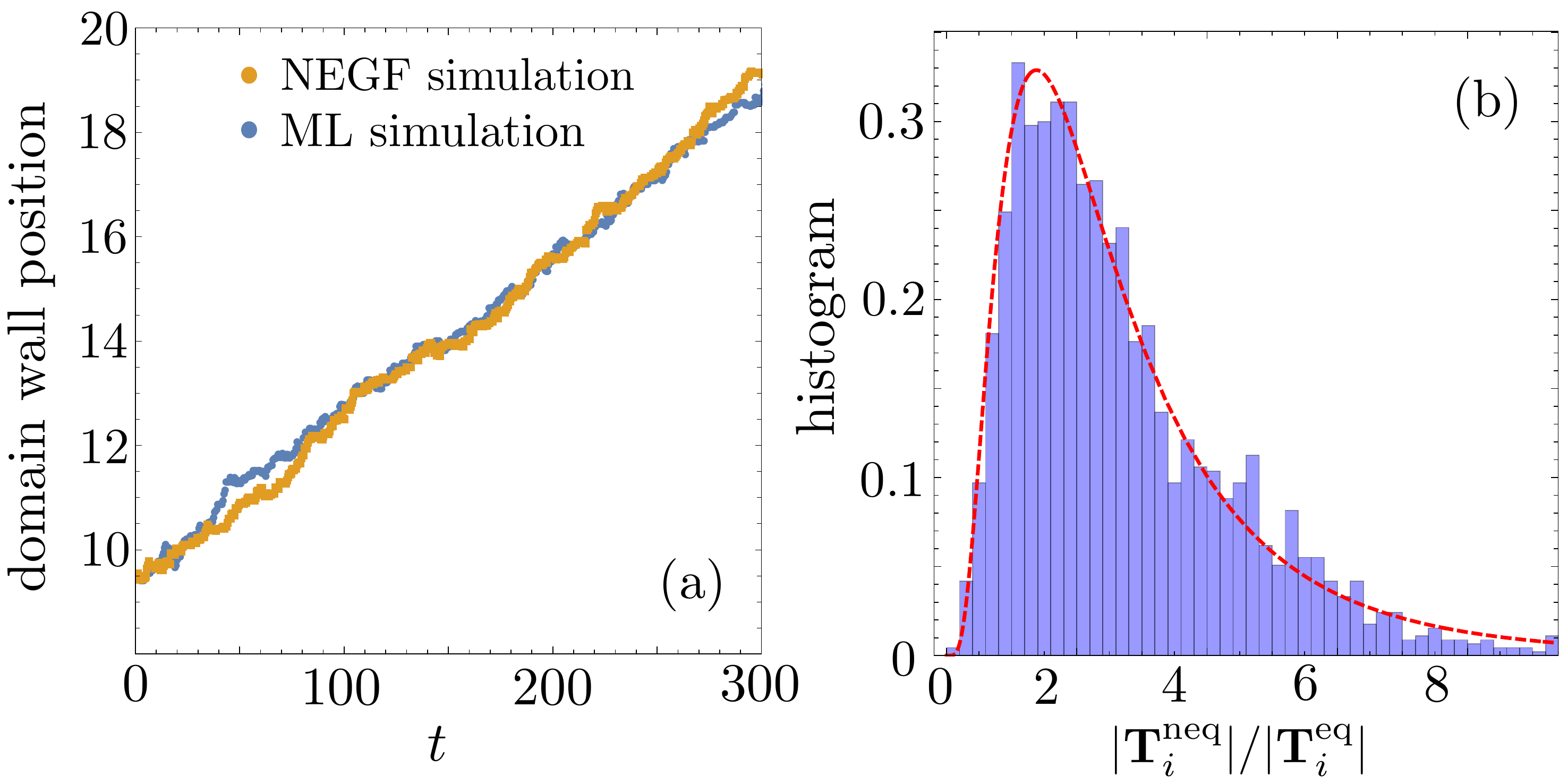}
\caption{(a) Average position of FM-AFM domain, obtained from NEGF-LLG and ML-LLG simulations, as a function of time during the voltage-driven insulator to metal transition of the s-d model.  (b) histogram of the ratio $|\mathbf T^{\rm neq}_i|/ |\mathbf T^{\rm eq}_i|$ predicted by the NN model for the simulation of domain-wall propagation. Here $\mathbf T^{\rm eq}_i = \mathbf S_i \times \partial \mathcal{E} / \partial \mathbf S_i$ is the quasi-equilibrium torque, and $\mathbf T^{\rm neq}_i =  \mathbf S_i \times (\mathbf S_i \times \partial \mathcal{G} / \partial \mathbf S_i)$ is the non-equilibrium torque in the generalized LLG Eq.~(\ref{eq:LL2}). 
\label{fig:benchmark} 
}
\end{figure}

The spatio-temporal dynamics of this insulator-to-metal phase transformation has recently been systematically studied based on the NEGF-LLG simulations~\cite{chern21}.  The kinetics of the nonequilibrium phase transition described above is essentially governed by the propagation of the FM-AFM domain walls. Fig.~\ref{fig:domain} shows the propagation of domain walls obtained from the NEGF-LLG as well as the ML-LLG simulations on a $30\times 24$ square lattice. The same initial state with a well-developed FM-AFM domain wall was used for both simulations.  The domain-wall position averaged over the transverse $y$-direction is plotted in Fig.~\ref{fig:benchmark}(a) as a function of time for both the NEGF and ML-LLD simulations.  Importantly, these comparisons demonstrate that our NN model not only is capable of predicting the nonequilibrium forces, but also can be used to accurately reproduce the nonequilibrium dynamical process of voltage-driven domain-wall propagation. The small discrepancy can be attributed to the random Langevin noise in the NEGF-LLG simulation and the force prediction error of the ML model.

A useful by-product of our NN model is the partition of the electron forces in Eq.~(\ref{eq:H_NEGF}), into the quasi-equilibrium $\mathbf h^{\rm eq}$ and nonequilibrium $\mathbf h^{\rm neq}$ components. As demonstrated in Fig.~\ref{fig:decomposition}(c), the quasi-equilibrium torque $\mathbf T^{\rm eq}_i =  \mathbf h^{\rm eq} \times \mathbf S_i$ is responsible for the precession motion of spins along contours of constant energy~$\mathcal{E}$. On the other hand, the nonequilibrium torque $\mathbf T^{\rm neq}_i = \mathbf h_i^{\rm neq} \times \mathbf S_i$ often points to a direction away from the (quasi) equilibrium field, thus acting similar to the so-called anti-damping torques~\cite{brataas12,ralph08,salahuddin08}.  Fig.~\ref{fig:benchmark}(b) shows a histogram of the ratio of these two torque components for spins in the vicinity of the AFM-FM domain walls. As expected, the driving force of the domain-wall propagation is dominated by the nonequilibrium forces.

To summarize, we have developed a new potential theory of the nonequilibrium  forces for the Landau-Lifshitz spin dynamics. This formulation allows us to generalize the Behler-Parrinello scheme, which is fundamental to most ML potentials for equilibrium quantum molecular dynamics and spin dynamics methods, to the modeling the electronic forces in out-of-equilibrium systems. We demonstrate our approach by developing a neural network model that successfully predicts the electronic forces computed from the nonequilibrium Green's function method for a driven s-d model. Integrating the ML potential with the LLG simulations, we further show that the nonequilibrium process of voltage-driven domain-wall motion can be accurately reproduced by the trained neural network model. 

Our work marks the first crucial step towards the multi-scale dynamical modeling of out-of-equilibrium quantum materials based on ML methods.  In addition to the dynamical modeling of voltage-driven insulator-to-metal transition as demonstrated in this work, our ML formulation is also capable of describing the current-induced spin-transfer torques, which play a crucial role in the field of spintronics. The ML potential proposed in this work thus lays the groundwork for large-scale Landau-Lifshitz dynamics simulations with the desired quantum accuracy for the nonequilibrium electron forces, such as the spin-transfer torques. Finally, we note that our results cannot be directly applied to the atomic forces. Much work is required to develop a similar framework for the ML-potential for nonequilibrium quantum MD simulations.

\begin{acknowledgements}
{\em Acknowledgements}. The authors thank Sheng Zhang and Avik Ghosh for useful discussions. This work was supported by the US Department of Energy Basic Energy Sciences under Award No. DE-SC0020330. The authors also acknowledge the support of Research Computing at the University of Virginia.
\end{acknowledgements}

\end{document}